\documentstyle[12pt]{article}
\newcommand {\m}{\mu}
\newcommand {\n}{\nu}
\newcommand {\pl}{\partial}

\newcommand {\al}{\alpha}
\newcommand {\be}{\beta}
\newcommand {\ga}{\gamma}

\newcommand {\ka}{\kappa}
\newcommand {\la}{\lambda}

\newcommand {\si}{\sigma}
\newcommand {\om}{\omega}

\newcommand {\ep}{\epsilon}
\newcommand {\vep}{\varepsilon}

\newcommand {\na}{\nabla}
\newcommand {\del}  {\delta}

\newcommand {\mn}{{\mu\nu}}
\newcommand {\ls}   {{\lambda\sigma}}
\newcommand {\ab}   {{\alpha\beta}}
\newcommand {\half}{ {\frac{1}{2}} }

\newcommand {\sqg} {\sqrt{g}}

\newcommand {\Lcal}{{\cal L}}

\newcommand {\G}[2] { {\Gamma^{#1}_{#2}}  }

\newcommand {\change} {\leftrightarrow}
\newcommand {\pr}   {{\quad .}}
\newcommand {\com}  {{\quad ,}}
\newcommand {\q}    {\quad}
\newcommand {\qq}   {\quad\quad}
\newcommand {\qqq}   {\quad\quad\quad}
\newcommand {\qqqq}   {\quad\quad\quad\quad}
\newcommand {\qqqqq}   {\quad\quad\quad\quad\quad}

\newcommand {\nl}    {\newline}
\newcommand {\vs}[1]  { \vspace*{#1 cm} }

\newcounter{eq}
\newcounter{sc}
\newcommand {\addeq}{\addtocounter{eq}{1}}
\newcommand {\addsc}{\addtocounter{sc}{1}}
\newcommand {\eqn}  {\mbox{(\thesc.\theeq)}}
\newcommand {\eqna} {\mbox{(\thesc.\theeq a)}}
\newcommand {\eqnb} {\mbox{(\thesc.\theeq b)}}
\newcommand {\eqnc} {\mbox{(\thesc.\theeq c)}}
\newcommand {\eqnd} {\mbox{(\thesc.\theeq d)}}
\newcommand {\eqne} {\mbox{(\thesc.\theeq e)}}
\newcommand {\eqnf} {\mbox{(\thesc.\theeq f)}}

\newcommand {\MPL}  {Mod.Phys.Lett.}
\newcommand {\NP}   {Nucl.Phys.}

\newcommand {\PR}   {Phys.Rev.}

\newcommand {\cen}[1]    { \multicolumn{1}{c|}{#1} }
\newcommand {\cena}[1]    { \multicolumn{1}{|c|}{#1} }
\begin{document}
\title{   Graphical Representation of Invariants and
          Covariants in General Relativity
          \thanks{US-93-06,
\ The content of this paper was in part reported at  the annual meetings of
the Physical Society of Japan in the autumn of 1992 and in the spring of
                                   1993.}                                 }
\author{  Shoichi ICHINOSE \\
          Department of Physics, University of Shizuoka,\\
          Yada 52-1, Shizuoka 422, Japan
          \thanks{
          E-mail address:                     \nl
          ICHINOSE@U-SHIZUOKA-KEN.AC.JP(JUNET,Preferable address)\nl
          ICHINOSE@JPNYITP(BITNET)
                  }
          }
\date{  July  1993  }
\maketitle
\setlength{\baselineskip}{0.8cm}
\begin{abstract}
We present a grapical way to describe invariants and covariants
in the (4 dim) general relativity. This makes us free from the complexity
of suffixes .
Two new off-shell relations between (mass)$\ast\ast 6$\ invariants
are obtained.
These are important for 2-loop off-shell calculation in the
perturbative quantum gravity.
We list up all independent invarians with dimensions of
(mass)$\ast\ast 4$\ and (mass)$\ast\ast 6$.
Furthermore  the explicit form of 6 dim Gauss-Bonnet
identity is obtained.
\end{abstract}
\vspace{2 cm}
\setcounter{sc}{0}
\addsc
\setcounter{eq}{0}
\begin{center}
{\Large\bf \S 1. Introduction}
\end{center}
In relation to the developement of the unified theory and the
investigation of the initial stage of the universe, the physical
importance of the quantum gravity grows more and more. These several
years the statistical
aspect of the (euclidean) quantum gravity (critical dimension,\ etc.)
has been vigorously investigated
and has been clarified using lower-dimensional models.
The renormalizability problem,however,does not go beyond the pionering
work by 'tHooft and Veltman[TV]. The problem seems crucial for approaching
the quantum gravity perturbatively[I1,I2].
In such research we must treat invariants with higher
mass-dimensions: (mass)$^{2+2n}$ for n-loop perturbation-order
( in the ordinary gauge).
In this circumstance it seems important to develope a systematic and
convenient way to deal with those invariants.
We present a graphical way, which enables us visually discriminate among
various invariants.
It makes us free from complexity of suffix-contraction
and reduces the algebraic computation considerably.
Mathematically this is a representation of invariants
(covariants) in terms of graphs.

As is well known, the symmetry of Rieman curvature tensor is not
so simple (see Sect.2). So far there has been
no systematic and practically-
useful method to construct independent invariants [P,G,VW].
In particular the relations between invariants at the off-shell level
( ,that is,with no use of a field equation )
have not been so much investigated.
It is known,however, that those relations are so important to envestigate
the quantum gravity [VW,I1,I2].
We present a new method using
a graph and derive all off-shell relations among the (mass)$\ast\ast 4$\
invariants and among the (mass)$\ast\ast 6$\ invariants .

In Sect.2 we list all well-known symmetries and relations of  invariants
and covariants in the general relativity. A graphical represatation of
Riemann tensor and the covariant derivative is introduced in Sect.3.
Three rules, which are basic relations  in the graphical calculation,
are presented. In Sect.4 , all invariants with dimensiom of (mass)$\ast\ast
 4$
and with dimension of (mass)$\ast\ast 6$ are expressed graphically.
All listed graphs are classified in Sect.5.
We reduce ,in Sect.6,
the number of (mass)$\ast\ast 6$\ invariants to 8 by using the three
rules above and two new off-shell relations.
In Sect.7 we comment on the independence of finally listed invariants
in Sect.6. Furthermore the explicit form of Gauss-Bonett identity in
6 dim space-time is obtained.
\vspace{2 cm}
\addsc
\setcounter{eq}{0}
\begin{center}
{\Large\bf \S 2. Preliminaries  }
\end{center}

Before the main text,
we summarize the present notation and list up all well-known symmetry
properties of covariants.
\begin{flushleft}
1.\  Notation
\end{flushleft}
\addeq\begin{eqnarray*}
&  R^\mu_{~\nu\ab}=\pl_\al\G\mu{\nu\be} +\G\mu{\al\ga}\G\ga{\nu\be}
                                     -(\al\change\be)\ \ ,\ \
 \G\al\mn =\half g^\ab (\pl_\nu g_{\mu\be}+\pl_\mu g_{\nu\be}
                                  -\pl_\be g_\mn )\ ,           &   \\
& R_\mn =R^\al_{~\mu\nu\al}\com \q R=R^{~\mu}_\mu\pr             & \eqn
\end{eqnarray*} 
 From these definitions, we know the mass dimension of the Rieman curvature
tensor , $[ R^\mu_{~\nu\ab} ]=(\mbox{mass})\ast\ast 2\ $,
\ under the definition
: $[g^\ab ]\equiv (\mbox{mass})\ast\ast 0,
\ [\pl_\al ]\equiv (\mbox{mass})\ast\ast 1 $.
\begin{flushleft}
2.\  Symmetry
\end{flushleft}
\addeq\begin{eqnarray*}
& R_{\mn\ls}=R_{\ls\mn}=-R_{\n\m\ls}=-R_{\mn\si\la}\com  &\eqna \\
& R_{\mn\ls}+R_{\n\la\m\si}+R_{\la\m\n\si}=0\com          &\eqnb \\
& R_\mn=R_{\n\m}\com                                         &\eqnc \\
& \na_\del R_{\ab\mn}+\na_\mu R_{\ab\n\del}+\na_\n R_{\ab\del\m}=0
                          \ (\mbox{Bianchi identity})\com    &\eqnd \\
& \na^\al R_{\mn\la\al}=\na_\m R_{\n\la}-\na_\n R_{\m\la}\com
                                                             &\eqne \\
& \na^\m (R_\mn -\half g_\mn R)=0\com                        &\eqnf
\end{eqnarray*} 
In addition to the above symmetries which are derived by the manipulation
of local quantites, there exist one relation ,between
(mass)$\ast\ast 4$ invariants,
which is related to a global (topological) quantity.
\begin{flushleft}
3.\ Gauss-Bonnet identity(in 4 dim)
\end{flushleft}
\addeq\begin{eqnarray*}
& \int d^4x\sqg R_{\mn\ab}R_{\ls\ga\del}\vep^{\mn\ls}\vep^{\ab\ga\del}=& \\
& 4\ \int d^4 x\sqg (\ R^2+R_{\mn\ab}R^{\mn\ab}-4R_\mn R^\mn\ )=
                    \mbox{topological invariant}\com\qq   &  \eqn
\end{eqnarray*} 
where $\vep^{\mn\ls}$ \ is the totally antisymmetric tensors.
\begin{flushleft}
4.\ covariant ingredients
\end{flushleft}
In Table 1, we list up all covariant ingrediants that make up
invariants.
\vspace{1 cm}

\begin{tabular}{|c|c|c|c|c|}
\hline
Dim $\backslash$ Suffix \#
              &  $0$     &   $1$   &   $2$     &    $4$           \\
\hline
 (mass)$\ast\ast 0$   &          &         &  $g_\mn$  & $\vep_{\mn\ls}$  \\
\hline
 (mass)$\ast\ast 1$   &          & $\na_\m$ &           &                 \\
\hline
 (mass)$\ast\ast 2$   &   $R$    &         &  $R_\mn$   &  $R_{\mn\ls}$   \\
\hline
\multicolumn{5}{c}{}                  \\
\multicolumn{5}{c}{Table 1}
\end{tabular}
\vspace{2 cm}
\addsc
\setcounter{eq}{0}
\begin{center}
{\Large\bf \S 3. A Graphical Way to Represent Covariants and Invariants }
\end{center}

We introduce a graphical representation for Riemann tensor, $R_{\mn\ls}$,
as in Fig.1.
\vs 6
\nl
Note the following items.
\begin{description}
\item[A1.] The graph is expressed by two kinds of lines ( dotted lines and
solid lines) and two { \em vertex points} which connect two different kinds
of lines.
We call the dotted line the {\em suffix line} because it represents
the suffix flow.
The arrow indicates the direction of the suffix-flow. The suffix which flows
in (out) is defined to be the left (right) suffix of each doublet ($\mn$) or
($\ls$) which appear in $R_{\mn\ls}$.
The solid line represents the Rieman tensor itself.
\item[A2.] We  need not distinguish between upper and lower suffixes
because we are interested only in invariants.
($R_{\m~\ls}^{~\n}\ ,R_\m^{~\n\ls}\ ,R^{\mn\ls}\ ,\cdots $ \ are also
represented by Fig.1.)
\item[A3.] We may write the graph in any way we like unless we change the
graph topologically and change the suffix flow( Fig.2a).
\item[A4.] We may freely move the vertex point unless it jumps another vertex
point( Fig.2b).
\item[A5.] Different lines
may freely cross  each other( Fig.2c, Fig.2d).
\end{description}
\ \vs {16}\nl

The graph satifies the symmetries (2.2a) which Riemann tensor $R_{\mn\ls}$
has(Fig.3).
\vs 9
\nl
We represent the contraction of two suffixes simply by connecting
them by a suffix line. Examples are given in
Fig.4a,b and c which express $R_{\m~,\ls}^{~\m}=0,\ R^\m_{~\n,\la\mu}=
R_{\nu\la}\ ,R^\n_{~\n}=R\ $ respectively.
\vs 9
\nl
We note the following facts.
\begin{description}
\item[B1.] From the definition of the arrow,  the graph changes its
sign under the change of the suffix flow( Fig.5a).
\item[B2.] Generally all suffix-lines are closed for invariants(Fig.4c).
Fig.4c, which represent $R$\ , is the simplest case of the graphical
represatation of invariants: the (mass)$\ast\ast 2$\ invariant .
\item[B3.] The arrow within a suffix line (closed or not-closed) may be
omitted
if even number of solid lines are connected to the line
because there exists no ambiguity in the direction of arrow for such a case
(Fig.4b,Fig.5b).
\item[B4.] Fig.4a says 'tadpole' graphs are prohibited.
\end{description}
\ \vs {14}
\nl

We can express the relation (2.2b) by a graphical rule :Rule 1(Fig.6).\nl
\vs 6
\nl
If we make use of the graphical representation, various relations
between products of Riemann tensors are easily  obtained. We
demonstrate here a simple example:
the relation of
$R_{\ga\del\tau\om}R_{\al~~~\be}^{~\tau\om}=\half R_{\ga\del\tau\om}
R^{\om\tau}_{~~~\ab}$\
(Fig.7) and its proof
using Rule 1(Fig.8). We call the graphical relation of Fig.7 \ Rule 2\
since it will be used as one of basic relations in the following.
\nl
\vs 5
\nl
\vs {13}
\nl
Similarly we can obtain another useful graphical rule
: Rule 3 (Fig.9a), using Rule 1.\nl
\vs {15}\nl
By contracting the suffixes $\m$ and $\la$ in Fig.9a, we obtain Fig.9b.
Furthermore Fig.9c is obtained by contracting $\n$ and $\si$ in Fig.9b.
We add here some notes.
\begin{description}
\item[C1.] Rule 2 \ is useful when we want to increase or decrease the number
of suffix-loops.
\item[C2.] Rule 3 \ is useful when we want to change the flow of a suffix
line.
\end{description}

In addition to the graphical representation for the curvature tensors,
we introduce that for the covariant derivative as
in Fig.10a($\na_\al R_{\mn\ls}$) and Fig.10b($\na_\be\na_\al R_{\mn\ls}$).
\vs 6\nl
In Fig.10a and b, the symbol $\na$\ represents the covariant derivative.
We must
depict it noting the following points.
\begin{description}
\item[D1.] The symbol $\na$\ must be written near enough a solid line to
show
clearly which Rieman tensor the derivative acts on.
\item[D2.] The order of the covariant derivatives is defined by;
 from the right to the left.
\end{description}
We can represent the Bianchi relations (2.2d),(2.2e) and (2.2f) as
in Fig.11a, Fig.11b and Fig.11c respectively.
\vs {12}\nl
\vspace{2 cm}
\addsc
\setcounter{eq}{0}
\begin{center}
{\Large\bf \S 4. Graphical Representation of Invariants
}
\end{center}

Now we list up all invariants with a fixed mass-dimension and represent
them graphically.
\begin{flushleft}
{\Large \S 4.1 Invariants with Dimension (Mass)$\ast\ast 4$}
\end{flushleft}
We can easily find the three invariants ,with the mass
dimension (mass)$\ast\ast 4$\ ,
as shown in
Fig.12a,b and c which represent $R^2,R_\mn R^\mn $\  and
$R_{\mn\ls}R^{\mn\ls}$\ respectively.
\vs 8\nl
Note that we have the relation Fig.9c and
we need not consider the total derivative :$\na_\mu \na^\mu R$\ .

It is well known, in 4 dim quantum gravity, that there exists one
relation between three invariants above: Gauss-Bonnet identity (2.3).
Here we note that the integrant of (2.3),
$R_{\mn\ab}R_{\ls\ga\del}\vep^{\mn\ls}\vep^{\ab\ga\del}$\ ,
can be expressed graphically as in Fig.13.
\vs 5\nl
One part (,say, the upper part) of the contracted suffixes
( the upper and lower suffixes ) are explicitly written in Fig.13
in order to specify which suffixes are anti-symmetrized.
We will compare Fig.13 with its 6-dim counter-part in sect6.

Therefore we can take two out of three invariants above,
as (locally) independent invariants;say,
\addeq\begin{eqnarray*}
R^2\qq,\q R_\mn R^\mn \qqqqq\qqqq \eqn
\end{eqnarray*} 
We will comment on their independence in sect.6.
\begin{flushleft}
{\Large \S 4.2 Invariants with Dimension (Mass)$\ast\ast 6$}
\end{flushleft}
There exist two types of invariants with the mass dimension (mass)$\ast\ast
 6$
: \ $\na R\times \na R$ and
\ $R\times R\times R$.
\begin{flushleft}
(i)\ $\na R\times \na R$\ (Fig.14a,b)
\end{flushleft}
\ \vs 6
\nl
Fig.14a and b represent $O_1\equiv \na^\mu R\cdot \na_\mu R$\ and
$O_2\equiv \na^\mu R_\ab\cdot \na_\mu R^\ab$\ respectively.
\begin{flushleft}
{(ii)\ $ R\times R\times R$}
\end{flushleft}
Products of three curvature tensors can be listed as in Fig.15a-f
and Fig.16a-f.
The number (0 or 2 or 4) indicates that of suffixes each curvature tensor
has. ( 0,2 and 4 imply  the scalar curvature  $R$\ , \
Ricci curvature tensor $R_\mn$\ ,
and Rieman curvature tensor $R_{\mn\ls}$ respectively.
We call here all these quantites curvature tensors.)\nl
\vs {15}\nl
Fig.15a,b,c,d,e and f represent
$P_1=RRR,P_2=RR_\mn R^\mn,P_3=RR_{\mn\ls}R^{\mn\ls},
P_4=R_\mn R^{\n\la}R_\la^{~\mu},P_5=R_{\mn\ls}R^{\mu\la}R^{\nu\si},
P_6=R_{\mn\ls}R_\tau^{~\nu\ls}R^{\mu\tau}$,respectively.

Note that \{ $O_i$ \} and \{ $P_i$ \} are
those invariants which vanish on shell,
$R_\mn =0$\ :the field equation of  Einstein-Hilbert lagrangian
$\Lcal =(1/\ka)\sqg R$.
The remaning kind are (4,4,4):\ product of three Rieman
curvature tensors
which does not vanish on shell. We can list them up as in Fig.16a-f.\nl
\vs {15}\nl
Fig.16a,b,c,d,e and f \ represent
$A_1=R_{\mn\ls}R^{\si\la}_{~~~\tau\om}R^{\om\tau\n\m}\ ,
 B_1=R_{\mn\tau\si}R^{\n~~~\tau}_{~\la\om}R^{\la\mu\si\om}\ ,
 B_2=R_{\mn\om\tau}R_{\ls}^{~~~\tau\om}R^{\si\m\n\la}\ ,
 B_3=R_{\mn\om\tau}R^{\n~\tau}_{~\la~~\si}R^{\la\m\si\om}\ ,
 C_1=R_{\mn\si\tau}R^{\n~~~\m}_{~\la\om}R^{\ls\tau\om}\ ,$  and
$C_2=R_{\mn\si\tau}R^{\n~\tau}_{~\la~~\om}R^{\ls\om\m}\ ,$
respectively.

We will derive all relations between \{ $O_i,P_i,A_1,B_i,C_i$ \} in
Sect.6.
( Here we list the relation between invariants defined in other references
and those defined in the present paper. As for [VW] ,
$A_1^{VW}=A_1\ ,A_2^{VW}=-B_1\ ,A_3^{VW}=C_2\ .$\ As for [I1],
$O_1^I=P_1\ ,O_2^I=P_2\ ,O_3^I=P_4\ ,O_4^I=P_5\ ,O_5^I=P_3\ ,
O_6^I=P_6\ ,O_7^I=-A_1\ ,O_8^I=O_1\ ,O_9^I=O_2\ .$\ )
\vspace{2 cm}
\addsc
\setcounter{eq}{0}
\begin{center}
{\Large\bf \S 5. Classification of Graphs}
\end{center}
Here we classify all graphs ,which have appeared in previous sections,
in order to confirm no missing graphs. Let us consider a graph with
$N_R$ solid lines and $N_E$ external suffix lines. Its mass-dimension
is (mass)$\ast\ast$(2$N_R$). The number of vertices is $V=2N_R$\ .\
The number of internal suffix lines,$N_I$, equals the number of suffix-
contraction and is given by $N_I=2N_R-\half N_E$.
The number of total (internal and external) suffix lines is given by
$S=N_I+N_E=2N_R+\half N_E$. Particularly the relation
$3V=2(N_I+N_R)+N_E$\ holds true. Some simple examples are listed in
Table 2.
\vs 1

\begin{tabular}{|p{7cm}|p{1cm}|p{1cm}|p{1cm}|p{1cm}|p{1cm}|}
\hline
\cena{Graph}
  & \cen{$N_R$}  & \cen{$N_E$}  & \cen{$V$}    & \cen{$N_I$}  &  \cen{$S$} \\
\hline
\vs 4  & \cen{1} & \cen{4}      & \cen{2}      & \cen{0}      & \cen{4} \\
\hline
\vs 4  & \cen{1} & \cen{2}      & \cen{2}      & \cen{1}      & \cen{3} \\
\hline
\vs 4  & \cen{1} & \cen{0}      & \cen{2}      & \cen{2}      & \cen{2} \\
\hline
\multicolumn{6}{c}{}                                   \\
\multicolumn{6}{c}{Table 2}
\end{tabular}

\vs 1
\begin{flushleft}
1)\ $N_E=0$
\end{flushleft}
In this case the graph in our consideration is an invariant.
All suffix lines are internal and make up a number of closed-loops.
Let us define the number of those closed loops which have n vertices, $L_n$.
There cannot exist 'self-energy' graphs of purely suffix lines. This means
$L_0=0$. No tadpole graphs (B4 in Sect.3) means $L_1=0$. Then the following
relation holds true.
\addeq\begin{eqnarray*}
&N_I=2L_2+3L_3+4L_4+\cdots\com\q L_n:\mbox{non-negative integer} \qq& \eqn
\end{eqnarray*} 
We can use this formula to classify all invariants. Three cases
($N_R=1,2,3$)
are classified in Table 3 and corrsponding graphs in the previous sections
are given.
\vs 1

\begin{tabular}{|p{1cm}|p{1.5cm}|p{2.5cm}|p{5cm}|}
\hline
\cena{$N_R$} & \cen{$N_I=2N_R$} & \cen{$(L_n\neq 0)$}
                                            & \cen{Corresponding Graphs}  \\
\hline
\cena{1}     & \cen{2}          & \cen{$L_2=1$}      & \cen{Fig.4c}      \\
\hline
            &                  & \cen{$L_2=2$}
                                         & \cen{Fig.12a(disconnected),}  \\
\cena{2}     & \cen{4}          &                    & \cen{Fig.12c}     \\
\cline{3-4}
            &                  & \cen{$L_4=1$}      & \cen{Fig.12b}     \\
\hline
            &                  & \cen{$L_2=3$}
                                         & \cen{Fig.15a(disconnected),}  \\
            &                  &
                                         & \cen{Fig.15c(disconnected),}  \\
            &                  &                     & \cen{Fig.16a}    \\
\cline{3-4}
\cena{3}     & \cen{6}          & \cen{$L_3=2$}       & \cen{Fig.16b,
                                                            Fig.16d}  \\
\cline{3-4}
            &                  & \cen{$(L_2=1,L_4=1)$}
                                         & \cen{Fig.15b(disconnected),}  \\
            &                  &                     & \cen{Fig.15f,
                                                            Fig.16c}    \\
\cline{3-4}
            &                  & \cen{$L_6=1$}       & \cen{Fig.15d,
                                                            Fig.15e}    \\
            &                  &                     & \cen{Fig.16e,
                                                            Fig.16f}    \\
\hline
\multicolumn{4}{c}{}                                   \\
\multicolumn{4}{c}{Table 3}
\end{tabular}

\vs 1
\begin{flushleft}
2)\ $N_E\neq 0$
\end{flushleft}
The graph in our consideration is a covariant (or contravariant)
with $N_E$ suffixes. Among internal suffix lines, some of them make up
closed-loops. The other ones do not make up loops. They are linked,
by vertices,
directly to external suffix lines or to adjacent internal suffix
lines which are linked finally,through successive linkage ,to external
suffix lines. Let us define the number of such non-loop internal suffix
lines,
$K$\ . Then we have
\addeq\begin{eqnarray*}
&N_I=2L_2+3L_3+4L_4+\cdots+K\com\q
L_n:\mbox{non-negative integer} \qq& \eqn
\end{eqnarray*} 
We define the number of those sets of non-loop internal suffix lines
which are linked adjacently by n vertices, $K_{n-1}$, for $n\geq 2$.
Then we have
\addeq\begin{eqnarray*}
&K=K_1+2K_2+3K_3+\cdots\com\q
K_n:\mbox{non-negative integer} \qq& \eqn
\end{eqnarray*} 
Some simple examples are listed in Table 4. (5.2) and (5.3) are substituted
for (5.1) in the present case of $N_E\neq 0$.
\vs 1

\begin{tabular}{|p{5.5cm}|p{1cm}|p{1cm}|p{2cm}|p{2cm}|p{1cm}|p{1cm}|}
\hline
\cena{Graph} & \cen{$N_R$}  & \cen{$N_E$}  & \cen{$(L_n\neq 0)$}
                    & \cen{$(K_n\neq 0)$}     &  \cen{$N_I$} & \cen{$K$} \\
\hline
\vs 4       & \cen{2}      & \cen{2}      & \cen{$L_2=1$}
                    & \cen{$K_1=1$}           & \cen{3}      & \cen{1}  \\
\hline
\vs 4       & \cen{3}      & \cen{2}      & \cen{$L_3=1$}
                    & \cen{$K_2=1$}           & \cen{5}      & \cen{2}  \\
\hline
\vs 4       & \cen{4}      & \cen{2}      & \cen{$L_4=1$}
                    & \cen{$K_3=1$}           & \cen{7}      & \cen{3}  \\
\hline
\multicolumn{6}{c}{}                                   \\
\multicolumn{6}{c}{Table 4}
\end{tabular}

\vs 1
As an example, let us classify the case $N_R=2,N_E=4$. (5.2) and (5.3) are
written as
\addeq\begin{eqnarray*}
&2=2L_2+3L_3+4L_4+\cdots+K\com\q & \\
&K=K_1+2K_2+3K_3+\cdots\com\q & \eqn
\end{eqnarray*} 
All possible cases for the choice of $(L_n)$ and $(K_n)$ are listed up in
Table 5.
\vs 1

\begin{tabular}{|p{0.5cm}|p{0.8cm}|p{0.8cm}|p{0.5cm}|p{4cm}|p{4cm}|}
\hline
          & \cen{$(L_n$}
                   & \cen{$(K_n$}
                                &          & \cen{Connected}
                                                  & \cen{Disconnected}  \\
\cena{$K$} & \cen{$\neq 0)$}
                   & \cen{$\neq 0)$}
                                & \cen{$N_E'$}
                                           & \cen{ Graphs}
                                                  & \cen{ Graphs}\\
\hline
\cena{0}   & \cen{$L_2=1$}
                   &            & \cen{4}  &      & \vs 4 \\
\hline
          &        & \cen{$K_2=1$}
                                & \cen{2}  & \vs 4 & \cen{No Graph} \\
\cline{3-6}
\cena{2}   &        & \cen{$K_1=2$}
                                & \cen{0}  &       & \vs 7           \\
\hline
\multicolumn{6}{c}{}                                   \\
\multicolumn{6}{c}{Table 5}
\end{tabular}

\vs 1
In Table 5, $N_E'$ is the number of those external suffix lines which are
not linked to internal suffix lines. $N_E'$ is given by
$N_E'=N_E-2(K_1+K_2+\cdots)$. The connected graphs with $N_E'=0$
correspond to 'one-particle irreducible' graphs. The all connected graphs
in Table 5 appear in Fig.7(Rule 2) and in Fig.9a(Rule 3).
\vspace{2 cm}
\addsc
\setcounter{eq}{0}
\begin{center}
{\Large\bf \S 6. Application of Graphical Representation}
\end{center}
We can prove the following relations among \{ $A_1,B_i,C_i$\}
in the  way similar to the proof of Rule 2 (Fig.8) in Sect.3. We list
the relations with the names of rules which are neccesary for
their proofs.
\addeq\begin{eqnarray*}
& B_2=\half A_1\q (\mbox{Rule 2})\com B_3=C_2 - C_1\q (\mbox{Rule 3})
                        =- B_1\ (\mbox{relation below})\com   &      \\
& C_1=\frac {1}{4} A_1\q (\mbox{Rule 2, Rule 2})\com        & \eqn \\
& C_2=C_1-B_1\q (\mbox{Rule 1})=\frac{1}{4} A_1-B_1
\ (\mbox{previous relation})\ .  &
\end{eqnarray*} 
Therefore we see all invariants of the kind (4,4,4)
are described only by $A_1$ and
$B_1$\ [VW].

There exist a different kind of relations which come not from the rules
in sect.3 but from the dimension of the space-time:\ 4\ .
\begin{flushleft}
Off-Shell Relation 1
\end{flushleft}
Let us consider the  identity of
Fig.17. This idea was explicitly noticed in [GS].
\vs 6\nl
Note that the identity Fig.17 holds true because each Greek suffix runs
from 0 to 3 (or from 1 to 4 for the Eucledian gravity) in 4 dim space-time.
This identity can be written as (by use of a computer)
\addeq\begin{eqnarray*}
&-P_2+\half P_3+2P_4-4P_5-5P_6+A_1-2B_1=0\pr\qqq  \eqn&
\end{eqnarray*} 
The on-shell case of (6.2), $A_1=+2B_1 $\ , was obtained in [VW]
by use of the spinor formalism.
\begin{flushleft}
Off-Shell Relation 2
\end{flushleft}
Similarly we can consider the identity of Fig.18.
\vs 5\nl
This identity can be written as (by use of a computer)
\addeq\begin{eqnarray*}
I\equiv 8(-P_1 +12 P_2-3 P_3 -16 P_4 +24 P_5 +24 P_6 -4 A_1 +8B_1) =0\pr\q
 \eqn
\end{eqnarray*} 
The on-shell case of (6.3) again gives
$A_1=+2B_1 $\ . At the off-shell level,however, (6.2) and (6.3) are
independent relations.
\begin{flushleft}
Off-Shell Relation 3
\end{flushleft}
 From the relations (6.2) and (6.3), we obtain the relation between
$\{ P_i \}$ as follows.
\addeq\begin{eqnarray*}
& - P_1\ +8 P_2\ - P_3\ -8 P_4\
+8 P_5\ +4 P_6=0\pr\q \eqn  &
\end{eqnarray*} 
This relation can also be directly derived by the identity of Fig.19.
\vs 5\nl
The independent relations are two out of the above three relations
(6.2),(6.3) and (6.4). We have checked any other choice of
anti-symmetrized suffixes and a starting graph, in the above
procedure, does not lead to another independent relation.

Therefore we can list up 8 (\ =2($O_i$)+6($P_i$)+2($A_1,B_1$)\ $-$ 2
(off-shell relations)\ )
independent
(mass)$\ast\ast 6$ invariants ,say, as follows.
\footnote{
In [GS], 9 terms are  listed. They missed the relation (6.4).
          }
\addeq\begin{eqnarray*}
& O_1\com\q O_2\com\q P_1\com\q P_2\com\q P_3\com\q P_4\com\q
P_5\com\q A_1\pr\q \eqn
\end{eqnarray*} 
As for their independence we make comment in Sect.7.
The importance of $A_1$\ ,which is the unique ,in (6.5), non-vanishing
term on shell, was first pointed out in [K].
\vspace{2 cm}
\addsc
\setcounter{eq}{0}
\begin{center}
{\Large\bf \S 7. Independence of Invariants , 6 Dim Gauss-Bonett Identity
and Conclusion}
\end{center}

The final proof of the independence of the listed invariants (4.1) and
(6.5) can  be done by the weak-gravity expansion approach. Let us introduce
the 'linear field' $h_\mn$\ instead of the metric $g_\mn$ by considering
the case of weak gravity :
\addeq\begin{eqnarray*}
g_\mn =\del_\mn +h_\mn\com |h_\mn|\ll 1\com\qqqq   \eqn
\end{eqnarray*} 
where $\del_\mn$ is the Minkowski metric (the flat space-time).
Each invariants can be expanded around the flat metric and can be
expressed by the infinite series of powers of $h_\mn$(with derivatives).
We have checked
the independence of the listed invariants (4.1) and (6.5) by looking
at the first few orders of $h_\mn$ by use of a computer.

The procedure presented in this paper is valid for any space-time
dimension.
All relations in this paper, except (2.3)\ (Fig.13), (6.2)\ (Fig.17) ,
(6.3)\ (Fig.18) and (6.4)\ (Fig.19), are valid irrespective of the
space-time dimension. Particularly in 6 dim space-time,
we can obtain the explicit
form of  Gauss-Bonett identity as follows.
\addeq\begin{eqnarray*}
& \int d^6x\sqg R_{\mn\ab}R_{\ls\ga\del}R_{\tau\om\ep\theta}
\vep^{\mn\ls\tau\om}\vep^{\ab\ga\del\ep\theta}
=\int d^6x\sqg (\mbox{Left Hand Side of Fig.18})   &   \\
& =\int d^6x\sqg I =\mbox{topological invariant}\com   & \eqn
\end{eqnarray*} 
where $I$ has appeared in (6.3) and the above formula
is explicitly re-written as
\addeq\begin{eqnarray*}
& 8\times\int d^6x\sqg ( -R^3+12 RR_\mn R^\mn -3 RR_{\mn\ls}R^{\mn\ls}
                       -16 R_\mn R^{\nu\la}R_\la^{~\mu}    &    \\
& +24 R_{\mn\ls}R^{\mu\la}R^{\nu\si}+24 R_{\mn\ls}R_\tau^{~\nu\ls}R^{\mu\tau}
-4R_{\mn\ls}R^{\si\la}_{~~~\tau\om}R^{\om\tau\nu\mu}
+8R_{\mn\tau\si}R^{\nu~~~\tau}_{~\la\om}R^{\la\mu\si\om} \   )  & \\
& =\mbox{topological invariant.} \qqq\eqn                        &
\end{eqnarray*} 
It is known, in 6 dim quantum gravity, 1-loop counterterms are given
by (mass)$\ast\ast$6 invariants in the ordinary gauge[VW].
They are given by the linear combination of the 9
(\ =2($O_i$)+6($P_i$)+2($A_1,B_1$)\ $-$ 1
(Gauss-Bonett identity)\ ) independent invariants,say, as follows.
\addeq\begin{eqnarray*}
& O_1\com\q O_2\com\q P_1\com\q P_2\com\q P_3\com\q P_4\com\q
P_5\com\q P_6\com\q A_1\pr\q \eqn
\end{eqnarray*} 
Because $A_1$\ does not vanish on shell,
1-loop counterterms , in 6 dim quantum gravity,
do not trivially vanish on-shell. The importance
of this fact was stressed in [VW]. It must be compared with 4 dim case (4.1),
where 1-loop counterterms trivially vanish.
The relation (7.3) and the left-hand-side of Fig.18 are
the 6 dim counter-part of the relation (2.3) and Fig.13 in 4 dim.

Finally we comment on those invariants which have not been considered
in the present paper.
\begin{enumerate}
\item
The extension to invariants with higher mass-dimension
(\ (mass)$\ast\ast 8$,$\cdots$\ ),in 4 space-time dimension,
is straightforward. They are important in higher-order perturbative
quantum gravity.
\item
Pseudoscalars: $\ep^{\mn\ls}R_{\mn\ab}R_\ls^{~~~\ab},
\ep^{\mn\ls}R_{\mn\ab}R_{\ls\ga\del}R^{\ab\ga\del},\cdots$. They does not
appear in the quantum effect in  Einstein-Hilbert theory. They might appear
in the unified theory including the gravitational and weak interactions.
\item
There are some non-local invariants which become local in a specific gauge.
The famous example is $R\frac{1}{\Delta} R$\ in the 2 dim gravity
,which is local with the conformal gauge.[PP]
It was exploited as the solvable model of 2 dim Eucledian quantum gravity.
Similar non-local invariants are discussed in 4  dim [AM].
At present,however, the role of those non-local invariants remain
obscure in 4 dim space-time.
\end{enumerate}

We recall that Feynman diagrams (graphs), which represent various scattering
processes of partricles, have been playing a very important role
in the practical and formal
development of QED, QCD and other renormalizable theories.
We hope the present approach will play an
analogous role  in the further
progress of the quantum gravity.

The results (6.2) and (6.3) and the proof of the independence
of terms in (4.1) and (6.5) by the weak-gravity expansion
are obtained by the algebraic software
FORM [V,VV].
\footnote{
The symbolic manipulation program FORM was written by J.A.M. Vermaseren.
Version 1.0 of the program and the manual are available via anonymous ftp
from
nikhef.nikhef.nl.
          }
\vspace{2 cm}
\begin{center}
{\Large\bf Acknowledgement }
\end{center}
The author thanks Prof. J.A.M. Vermaseren (NIKHEF-H,Amsterdam)
for kindly explaining the usage of
his original and excellent algebraic software FORM
in the spring of 1992 at KEK,Japan.
The author also thanks Dr. T.Watanabe (School of Food and Nutritional
Sciences, Univ. of Shizuoka) for kind help in drawing diagrams,
and Prof. N.Nakanishi (RIMS,Kyoto Univ.) for reading the manuscript
carefully.

\end{document}